\documentclass[preprint,secnumarabic,amssymb, showpacs, nobibnotes, aps, prd]{revtex4-1}

\usepackage{graphicx}
\usepackage{subfigure}
\usepackage{ae} 
\usepackage{float}  
\usepackage{amsmath}
\usepackage{color}

\begin{document}

\title{Generation of exchange magnons in thin ferromagnetic films by ultrashort acoustic pulses}

\author{V.Besse}
\author{A.V. Golov}
\author{V.S. Vlasov$^{1,2}$}
\email[]{vlasovvs@syktsu.ru}
\author{A. Alekhin$^1$}
\author{D. Kuzmin$^{3,4}$}
\author{I.V. Bychkov$^{3,4}$}
\author{L.N. Kotov$^3$}
\author{V.V. Temnov$^1$}
\email[]{vasily.temnov@univ-lemans.fr}

\affiliation{$^1$IMMM CNRS 6283, Le Mans Universit\'{e}, 72085 Le Mans cedex, France}
\affiliation{$^2$Syktyvkar State University named after Pitirim Sorokin, 167001, Syktyvkar, Russian Federation}
\affiliation{$^3$Chelyabinsk State University, 454001 Chelyabinsk, Russian Federation}
\affiliation{$^4$South Ural State University (National Research University),454080 Chelyabinsk, Russian Federation}

\begin{abstract}
We investigate generation of exchange magnons by ultrashort, picosecond acoustic pulses propagating through ferromagnetic thin films. Using the Landau-Lifshitz-Gilbert equations we derive the dispersion relation for exchange magnons for an external magnetic field tilted with respect to the film normal. Decomposing the solution in a series of standing spin wave modes, we derive a system of ordinary differential equations and driven harmonic oscillator equations describing the dynamics of individual magnon mode. The external magnetoelastic driving force is given by the time-dependent spatial Fourier components of acoustic strain pulses inside the layer. Dependencies of the magnon excitation efficiencies on the duration of the acoustic pulses and the external magnetic field highlight the role of acoustic bandwidth and phonon-magnon phase matching. Our simulations for ferromagnetic nickel evidence the possibility of ultrafast magneto-acoustic excitation of exchange magnons within the bandwidth of acoustic pulses in thin samples under conditions readily obtained in femtosecond pump-probe experiments.  
\end{abstract}

\pacs{Valid PACS appear here}


\maketitle

\section{Introduction}

The discovery of ultrafast laser-induced demagnetization in ferromagnetic nickel in 1996~\cite{beaurepaire1996ultrafast} opened new booming field of femtosecond laser manipulation of magnetization~\cite{koopmans2000ultrafast, van2002all, guidoni2002magneto, zhang2002coherent,vomir2005real,bigot2005ultrafast,kimel2005ultrafast,malinowski2008control,bigot2009coherent,bovensiepen2009femtomagnetism,radu2009laser,carpene2010ultrafast,boeglin2010distinguishing,Scherbakov2010,radu2011transient,rudolf2012ultrafast,bombeck2013magnetization,kim2012ultrafast,kim2015controlling,kim2017magnetization}. Depending on a mechanism, magnetization dynamics induced by a femtosecond laser pulse can be sorted into the following categories: (i) thermal effects~\cite{van2002all, carpene2010ultrafast}, (ii) magnetooptical effects~\cite{kirilyuk2010ultrafast}, (iii) magnetoacoustic effects~\cite{Scherbakov2010,Thevenard2010,kim2012ultrafast,bombeck2012excitation,temnov2012ultrafast,bombeck2013magnetization,januvsonis2016ultrafast,chang2017parametric} and (iv) spin transfer torque~\cite{nemec2012experimental, Schellekens2014, razdolski2017nanoscale,alekhin2017femtosecond}. 
Despite of the facts that the spin transfer torque is an efficient mechanism to drive coherent magnetization dynamics in ferromagnet through ultrashort pulses of spin-polarized electrons propagating in noble metals for several tens of nanometers before they lose their spin polarization~\cite{Alekhin2019}, acoustic pulses can propagate over much larger distances before they vanish. Picosecond acoustic pulses generated by femtosecond laser excitations ~\cite{ThomsenPRB86,temnov2012ultrafast,TemnovNatureComm2013,Temnov_2016} can drive the small-angle precession of ferromagnetic resonance (FMR) in various ferromagnetic samples  \cite{Scherbakov2010,Thevenard2010,kim2012ultrafast,bombeck2012excitation}. Experimental configurations using fs-laser excited periodic acoustic transients \cite{januvsonis2016ultrafast,chang2017parametric,kim2017magnetization} can be used to resonantly enhance the amplitude of FMR precession angle. Previous experimental work on ultrafast magnetoacoustics by Kim and co-workers~\cite{kim2015controlling,kim2017magnetization} demonstrated the possibility to control the ferromagnetic resonance (FMR)~\cite{Farle_1998} by a series of ultrashort, picosecond acoustic pulses. The theoretical treatment based on the phenomenological analysis of Landau-Lifshitz-Gilbert (LLG) equations accounting for some selected terms of the thermodynamic free energy density $F$ (the magnetocrystalline and magnetoelastic anisotropies, the demagnetization energy and the external magnetic field energy) was sufficient to interpret their experimental observations. Though, it did not allow to investigate the possibility to acoustically excite other elementary magnetic excitations such as exchange magnons \cite{van2002all}. 

While searching for fingerprints of elastic magnon excitation, Bombeck et al. \cite{bombeck2012excitation} reported on the magnetic-field dependence of two closely spaced frequencies in the magneto-optical response of an elastically driven 200~nm thick magnetic semiconductor (Ga,Mn)As. One of the modes was claimed to be the low-order exchange mode, but the theoretical intepretation rooted on a very specific spatial profiles of exchange magnons, strongly dependent on the unknown magnetic boundary conditions.  Kim and Bigot also observed the conspicuous beating of magneto-acoustic signals in a free-standing 300~nm thick ferromagnetic nickel film \cite{kim2017magnetization}, but explained the beating within the framework of the magnetoacoustic coherent control of FMR excitations, i.e. without involving any exchange magnons. Clearly, there is a need for a simple and transparent theory describing the magnetoelastic generation of exchange magnons using picosecond acoustic pulses.   

In this article we develop such magnetoelastic theory and predict the possibility of ultrafast magnetoelastic excitation of exchange magnons by ultrashort acoustic pulses in thin ferromagnetic films. This theory is based on the Landau-Lifshitz-Gilbert (LLG) equations where ultrashort acoustic strain pulses propagating through a ferromagnetic thin film modify its magnetoelastic energy and drive precessional dynamics of standing modes of exchange magnons, which can be also described by a simple equation of a driven harmonic oscillator. Being applied to a 30-nm nickel film excited by picosecond pulses of longitudinal acoustic phonons, our theory demonstrates the excitation of exchange magnons with frequencies within the bandwidth of the acoustic pulses.    

\section{Experimental geometry and main equations}

\begin{figure}[ht]
	\centering
    \includegraphics[width=1.0\columnwidth]{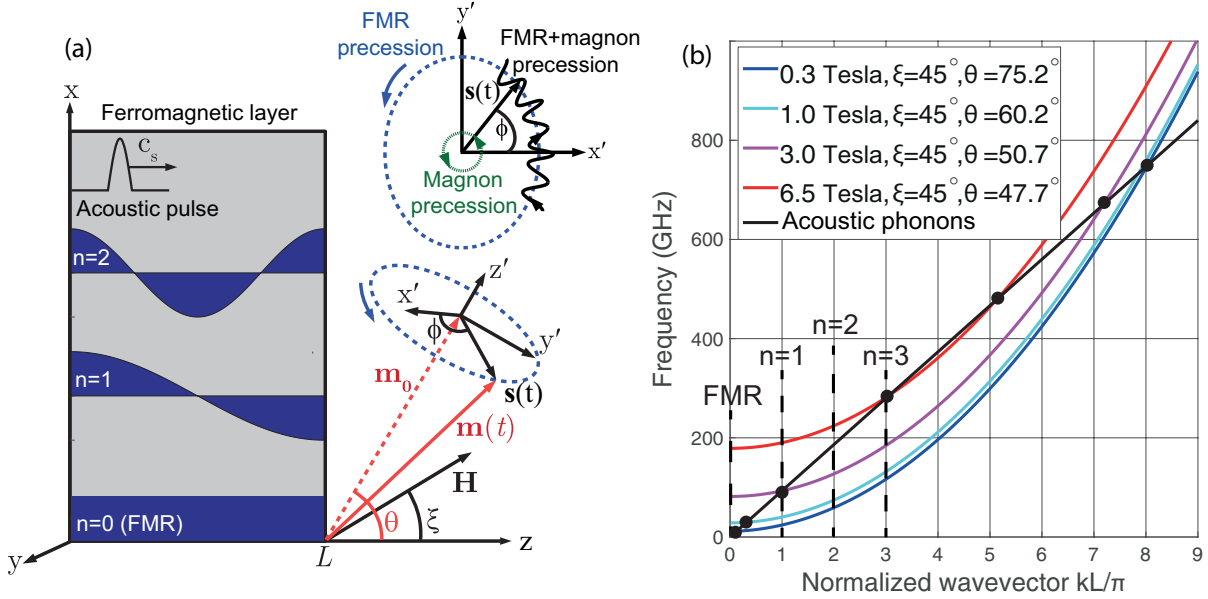}
	\caption{ (a) Picosecond pulses of longitudinal acoustic phonons propagating at the speed of sound $c_s$ through a ferromagnetic layer of thickness $L$ can excite simultaneously homogeneous (FMR, corresponding to the magnon mode with n$=$0) and non-uniform magnetization precession (magnon modes with n$>$0) around the equilibrium direction tilted by an angle $\theta$ with respect to the film normal. The external magnetic field is tilted by an angle $\xi$. (b) Magnon dispersion can be tuned by the amplitude of an external magnetic field.  The crossing points of magnon and phonon dispersions are marked with black dots.}
    \label{fig:schemeGeometry}
\end{figure}

Our numerical simulations are conducted for ferromagnetic nickel on which the pioneering ultrafast laser-induced demagnetization \cite{beaurepaire1996ultrafast} and the most recent magneto-acoustic experiments \cite{kim2012ultrafast,kim2015controlling,kim2017magnetization} have been performed by the group of Jean-Yves Bigot. We report a numerical study of exchange magnons in a 30-nm nickel film excited by picosecond acoustic pulses. Typical experimental configuration for magneto-acoustic \cite{kim2012ultrafast,kim2015controlling,kim2017magnetization} and magnonic \cite{Salikhov2019} measurements is presented in Fig.~\ref{fig:schemeGeometry} (a). It utilizes a rotating permanent magnet placed on the top of a ferromagnetic sample which allows to apply external magnetic field with a magnitude up to a few hundreds of millitesla at an arbitrary angle $\xi$ with respect to the surface normal. The equilibrium direction $\mathbf{m}_0$ of the magnetization vector $\mathbf{M}=M_0 \mathbf{m}_0$ is usually non-collinear with the external magnetic field due to the magnetic anisotropies and points at an angle $\theta$ with respect to the surface normal. Acoustic pulses with duration $\tau$ and spatial width $c_s\tau$ smaller than the sample thickness $L$ propagate through the film and locally alter the direction of the effective magnetic field $\mathbf{H}_{\mathrm{eff}}$ thereby driving precessional motion of the magnetization. The resulting magnetization dynamics can be represented as a sum of different magnon modes: homogeneous precession of the magnetization (FMR) and exchange-coupled non-uniform magnon modes.

\section{Magnetization dynamics excited by acoustic pulses}

The Landau-Lifschitz-Gilbert (LLG) equations
\begin{equation}
	\frac{\partial \mathbf{m}}{\partial t} = - \gamma \mu_0 \mathbf{m} \left( t, z \right) \times \mathbf{H}_{\mathrm{eff}} \left( t, z \right) + \alpha \mathbf{m} \left( t, z \right) \times \frac{\partial \mathbf{m}}{\partial t}
	\label{eq:LLG}
\end{equation}
represent the most common tool to model the spatio-temporal dynamics of the unit magnetization vector $\mathbf{m}\left(t,z\right)$ driven by the effective magnetic field $\mathbf{H}_{\mathrm{eff}}(z,t)$. Here, the $\alpha$ is the dimensionless phenomenological Gilbert damping parameter and  $\gamma$ demotes the gyromagnetic ratio. The effective magnetic field is a functional derivative of the free energy density
\begin{equation}
    \mathbf{H}_{eff} = - \frac{1}{\mu_0 M_0} \frac{\partial F}{\partial \mathbf{m}} + \frac{1}{M_0} \sum_{p=1}^{3} \frac{\partial}{\partial x_p} \frac{\partial U}{\partial \left(\frac{\partial  \mathbf{m}}{\partial x_p} \right)}\,,
\end{equation}
where $M_0$ is a saturation magnetization. For the purposes of this investigation we define the free density energy of a ferromagnetic thin film as a sum 
\begin{equation}
    F = F_{z} + F_{d} +  F_{ex} + F_{me}(t, z)\,,
    \label{eq:U}
\end{equation}
which includes the Zeeman conribution
\begin{equation}
    F_{z}=-\mu_0 M_0 \mathbf{m}\cdot\mathbf{H}\,
    \label{eq:Uz}
\end{equation}
with an external magnetic field $\mathbf{H}$, the demagnetizing field energy
\begin{equation}
    F_{d} = \frac{1}{2} \mu_0 M_0^2 \mathbf{m} \cdot \mathbf{N} \cdot \mathbf{m}\,
    \label{eq:Ude}
\end{equation}
determined by the demagnetization tensor   
\begin{equation}
    \mathbf{N} =  
   \left (\begin{array}{ccc}
      0 & 0 & 0 \\
      0 & 0 & 0 \\
      0 & 0 & 1 \\
   \end{array} \right )\,
    \label{eq:N}
\end{equation}
for a thin film geometry, the exchange energy
\begin{equation}
    F_{ex} = \frac{1}{2} M_0 \sum_{p=1}^3 D \left( \frac{\partial \mathbf{m}}{\partial x_p} \right)^2\,
    \label{eq:Uex}
\end{equation}
characterized by the exchange stiffness $D$, and the magnetoelastic energy
\begin{equation}
    F_{me}(t, z) = b_1m_z^2\epsilon_{zz}(z,t)\,,
    \label{eq:Ume}
\end{equation}
In the latter term, the magnetoelastic constant $b_1$ couples the normal magnetization component $m_z$ to the dynamic strain $\epsilon_{zz}(t, z)$ propagating in the $z$-direction (here our consideration is limited to a single non-zero strain component $\epsilon_{zz}$). While neglecting a weak dependence of the exchange stiffness $D$ on the applied strain, here we assume that the magnetoelastic interactions are driven solely by the last time-dependent magnetoelastic term $F_{me}(z,t)$ in Eq.~(\ref{eq:U}).

Given that the free energy density in Eq.~(\ref{eq:U}) represents a superposition of different terms, the effective magnetic field in Eq.~(\ref{eq:Heff}) also appears to be a sum of the corresponding contributions:
\begin{equation}
    \mathbf{H}_{\mathrm{eff}} = \mathbf{H} + \mathbf{H}_{\mathrm{d}} + \mathbf{H}_{\mathrm{ex}} + \mathbf{H}_{\mathrm{me}}(z,t)\,
    \label{eq:Heff}
\end{equation}
where $\mathbf{H}$ is the external magnetic field, $\mathbf{H}_{\mathrm{d}}$ is the demagnetization field, $\mathbf{H}_{\mathrm{ex}}$ is the exchange field and the time-dependent magnetoelastic field $\mathbf{H}_{\mathrm{me}}(t, z)$. After calculation of the functional derivatives from Eqs.~(\ref{eq:Uz}),~(\ref{eq:Ude}),~(\ref{eq:Uex}) and (\ref{eq:Ume}) and their sum, we obtain the total effective magnetic field: 
\begin{eqnarray}
    \label{eq:Heffx}
    H_\mathrm{eff,x} &=& D \frac{\partial^2 m_\mathrm{x}}{\partial z^2}+H \cos\xi\,,\\
    \label{eq:Heffy}
    H_\mathrm{eff,y} &=& D \frac{\partial^2 m_\mathrm{y}}{\partial z^2}\,,\\
    H_\mathrm{eff,z} &=& D \frac{\partial^2 m_z}{\partial z^2}+ H \sin\xi - M_0 \mathrm{m_\mathrm{z}} - \frac{2 b_1}{\mu_0 M_0} m_\mathrm{z} \epsilon_{\mathrm{zz}}(z,t)\,.
    \label{eq:Heffz}
\end{eqnarray}
Now, when the driving force for the LLG equations Eq.(\ref{eq:LLG}) is known, we first analyze the dissipation free case, i.e. $\alpha = 0$:
\begin{equation}
    \frac{\partial \mathbf{m}}{\partial t} = - \gamma \mu_0 \mathbf{m} \times \mathbf{H}_{eff}\,.
    \label{eq:LLG-2}
\end{equation}
Second, we introduce a small dynamic perturbation $|\mathbf{s}(z,t)|\ll |\mathbf{m}_0|$ of the magnetization vector oscillating around the equilibrium magnetization direction  $\mathbf{m}_0=(\sin\theta,0,\cos\theta)$:
\begin{eqnarray}
    m_x &=& \cos\theta+s_{x}(z,t)\,,
    \label{eq:mxd1}\\
    m_y &=& s_{y}(z,t)\,,
    \label{eq:myd1}\\
    m_z &=& \sin\theta+s_{z}(z,t)\,.
    \label{eq:mzd1}
\end{eqnarray}
By substituting Eqs.~(\ref{eq:mxd1}-\ref{eq:mzd1}) in Eqs.~(\ref{eq:Heffx}-\ref{eq:LLG-2}) and keeping only the linear terms in $s_{\mathrm{i}}$, we obtain the following system of differential equations:
\begin{eqnarray}
   \frac{1}{\gamma\mu_0}\frac{\partial {s}_{\mathrm{x}}}{\partial t}-D \cos \theta   \frac{\partial^2 {s}_{\mathrm{y}}}{\partial z^2} + (H \cos\xi - M_0 \cos\theta){s}_{\mathrm{y}}=
   \frac{2 b_1}{\mu_0M_0} \epsilon_{\mathrm{zz}}(t, z) {s}_{\mathrm{y}} \cos \theta\,,
    \label{eq:dmxdt1_0}
    \\
  \frac{1}{\gamma\mu_0}\frac{\partial {s}_{\mathrm{y}}}{\partial t}+D[\cos\theta  \frac{\partial^2 {s}_{\mathrm{x}}}{\partial z^2}
    - \sin\theta  \frac{\partial^2 {s}_{\mathrm{z}}}{\partial z^2}] - (H\cos\xi - M_0\cos\theta){s}_{\mathrm{x}}+    \nonumber\\ 
    +(H \sin\xi+M_0 \sin\theta){s}_{\mathrm{z}}=-\frac{2 b_1}{\mu_0M_0} \epsilon_{\mathrm{zz}}(z,t) \left( {s}_{\mathrm{x}} \cos \theta + {s}_{\mathrm{z}} \sin \theta+
    \sin \theta \cos \theta \right)\,,
    \label{eq:dmydt1_0}
    \\ 
   \frac{1}{\gamma\mu_0} \frac{\partial {s}_{\mathrm{z}}}{\partial t}+D \sin\theta \frac{\partial^2 {s}_{\mathrm{y}}}{\partial z^2} - (H\sin\xi){s}_{\mathrm{y}} =0\,.
    \label{eq:dmzdt1_0}
\end{eqnarray}
In real experiments, strains are usually small $\epsilon_{\mathrm{zz}}(z,t)<0.01$, which allows us also to neglect the mixed terms $\propto s_i\epsilon_{\mathrm{zz}}(z,t)$ in comparison with other terms linear in $s_i$ and $\epsilon_{\mathrm{zz}}$ and further simplify these equations to: 
\begin{eqnarray}
   \frac{1}{\gamma\mu_0}\frac{\partial {s}_{\mathrm{x}}}{\partial t}-D \cos \theta   \frac{\partial^2 {s}_{\mathrm{y}}}{\partial z^2} + (H \cos\xi - M_0 \cos\theta){s}_{\mathrm{y}}=0\,,
    \label{eq:dmxdt1}
    \\
  \frac{1}{\gamma\mu_0}\frac{\partial {s}_{\mathrm{y}}}{\partial t}+D[\cos\theta  \frac{\partial^2 {s}_{\mathrm{x}}}{\partial z^2}
    - \sin\theta  \frac{\partial^2 {s}_{\mathrm{z}}}{\partial z^2}] - (H\cos\xi - M_0\cos\theta){s}_{\mathrm{x}}+    \nonumber\\ 
    +(H \sin\xi+M_0 \sin\theta){s}_{\mathrm{z}}=-\frac{2 b_1\sin\theta\cos\theta}{\mu_0M_0}\epsilon_{\mathrm{zz}}(z,t)\,,
    \label{eq:dmydt1}
    \\ 
   \frac{1}{\gamma\mu_0} \frac{\partial {s}_{\mathrm{z}}}{\partial t}+D \sin\theta \frac{\partial^2 {s}_{\mathrm{y}}}{\partial z^2} - (H\sin\xi){s}_{\mathrm{y}} =0\,.
    \label{eq:dmzdt1}
\end{eqnarray}
The magnetoelastic term $\propto b_1\sin\theta\cos\theta\epsilon_{\mathrm{zz}}(z,t)$ on the right-hand side of Eq.~(\ref{eq:dmydt1}) drives the magnetization dynamics. It becomes zero for in-plane ($\theta=90^{\circ}$) or out-of-plane ($\theta=0$) static magnetization directions. This observation highlights the importance of a tilted magnetic configuration for magnetoelastic studies.     

Substituting the plane waves $s_i=c_i{\rm exp}(i\omega t-ikz)$ in Eq.~(\ref{eq:dmxdt1}, \ref{eq:dmydt1}, \ref{eq:dmzdt1}), where the aforementioned time-dependent magnetoelastic driving term is neglected, leads to the following secular equation 
\begin{equation}
\left |\begin{array}{ccc}
i\omega &  -A_{12}(k) & 0 \\
-A_{21}(k) & i\omega & -A_{23}(k)	 \\
 0 & -A_{32}(k)  & 	i\omega \\	
 \end{array} \right |=0\,,
 \label{eq:determ}
 \end{equation}
with coefficients $A_{ij}(k)$ defined as:
\begin{eqnarray}
    \label{eq:A12}
    A_{12}(k) &=& - \gamma \mu_0 \left[ \left( D k^2 - M_0 \right) \cos \theta + H \cos \xi \right]\\
    \label{eq:A21}
    A_{21}(k) &=& - A_{12}\\
    \label{eq:A23}
    A_{23}(k) &=& - \gamma \mu_0 \left[ \left( D k^2 + M_0 \right) \sin \theta + H \sin \xi \right]\\
    \label{eq:A32}
    A_{32}(k) &=& \gamma \mu_0 \left[ D k^2 \sin \theta + H \sin \xi \right]\,.
\end{eqnarray}
The secular equation provides the dispersion relation for the magnon modes at the frequency $\omega$ propagating in $z$-direction with the wave vector $k$:
\begin{equation}
    \omega(k) = \sqrt{-A_{12}A_{21}-A_{23}A_{32}}\,.
    \label{eq:mFourier}
\end{equation}
At $k=0$ and some particular orientations of the magnetic field, i.e. in-plane versus out-of-plane, this dispersion relation is reduced to well-known Kittel equations for the FMR frequency \cite{Farle_1998}. For large non-zero $k$, when exchange interactions dominate, the dispersion relation becomes quadratic $\omega(k)\sim Dk^2$. This dispersion relation can be tuned both by the amplitude and the direction of the external magnetic field, as shown in Fig.~\ref{fig:schemeGeometry}. As noted previously, the tilted orientation of the external magnetic field is crucial for magnetoelastic interactions. This is why we have conducted numerical simulations for an external magnetic field tilted by 45$^\circ$ with respect to the film normal and inspected the results as a function of the magnitude of the external magnetic field.

The dependence on the external magnetic field is shown in Fig.~\ref{fig:schemeGeometry}(b). For small magnetic field the magnon dispersion crosses the acoustic dispersion twice: at low frequency slightly above the FMR frequency and at very high frequency of order of several hundred GHz corresponding to large $k$-vectors. In strong magnetic fields the magnon dispersion is upshifted and both frequencies get closer and merge in a single point when the parabolic magnon dispersion touches the linear acoustic dispersion and the phase-matching is fulfilled over a wide range of frequencies. In case of monochromatic excitations the analysis of phase-matching conditions would be sufficient to predict the elastically driven magnetization dynamics. However, here we consider ultrashort acoustic pulse possessing very broad frequency spectrum and propagating through a ferromagnetic samples with a thickness of a few tens of nanometers, which requires the analysis beyond the phase-matching conditions.

In a ferromagnetic film with a finite thickness, only discrete number of magnonic modes are supported and survive on a larger time scale after the excitation. These modes are created by interference of two counter-propagating spin waves with wave vectors $\pm k_n$. Discretization of $k$-vector is determined by the boundary conditions for the dynamic magnetization at the interfaces between the magnetic layer and the adjacent material. Without the loss of generality we choose free boundary conditions 
\begin{equation}
    \left. \frac{\partial s_i}{ \partial z} \right|_{z = 0,L} =0\,,
    \label{eq:boundary}
\end{equation}
which result in cosine-like magnon eigenmodes $\propto\cos(k_nz)$ with $k_{\mathrm{n}} = \pi n/L$ and the dynamic magnetization  
\begin{equation}
	\mathbf{s}(z,t) = \sum_{n=0}^{N}{\mathbf{s}^{(n)}(t)}\cos k_nz\,
	\label{eq:Fourier}
\end{equation}
is represented as a sum over all contributing magnon modes; $N$ is the number of modes with non-zero amplitudes used in numerical calculations. We substitute this expression in the system of equations Eq.~(\ref{eq:dmxdt1}, \ref{eq:dmydt1}, \ref{eq:dmzdt1}), where we now keep the time-dependent magnetoelastic driving force represented in a form of the Fourier series. Due to orthogonality of the magnon eigenmodes, integration of the equations over the film thickness from 0 to $L$ leads to the system of {\it decoupled ordinary differential equations} for time-dependent amplitudes ${\mathbf{s}^{(n)}(t)}$ of all magnon modes:    
\begin{eqnarray}
    \label{eq:dmxdt2}
   \frac{ds^{(n)}_{\mathrm{x}}}{dt} &=&  A_{12}^{\left( n \right)}  s^{(n)}_{\mathrm{y}}\,,\\
    \label{eq:dmydt2}
    \frac{ds^{(n)}_{\mathrm{y}}}{dt} &=& A_{21}^{\left( n \right)} s^{(n)}_{\mathrm{x}} + A_{23}^{\left( n \right)} s^{(n)}_{\mathrm{z}} + \frac{2\gamma b_1\sin\theta\cos\theta}{M_0}\epsilon^{(n)}_{\mathrm{zz}}(t)\,,\\
    \label{eq:dmzdt2}
    \frac{ds^{(n)}_{\mathrm{z}}}{dt} &=& A_{32}^{\left( n \right)} s^{(n)}_{\mathrm{y}}\,,
\end{eqnarray}
with $A_{ij}^{\left( n \right)}=A_{ij}(k_n)$. 
The magnetization dynamics are driven by the time-dependent Fourier components of the elastic strain pulses   
\begin{equation}
\epsilon^{(n)}_{\mathrm{zz}}(t)=\frac{1}{L}\int_0^L\epsilon_{\mathrm{zz}}(z,t)\cos(k_nz)dz\,.
\end{equation}
The dynamic strain component $\epsilon_{\mathrm{zz,n}}$ acts on the $n_{th}$ magnon mode as an external driving force. Ultrashort, picosecond acoustic pulses with the length shorter than the film thickness produce a multitude of non-zero Fourier components.       
These Fourier components $\epsilon^{(n)}_{\mathrm{zz}}(t)$ are time-dependent: they change not only when an acoustic pulse enters or leaves a ferromagnetic film, but also when it propagates through the sample. The physical insight in the mechanisms of the magnetoelastic interactions in thick films can be obtained from the theoretical analysis of the integrals $\epsilon^{(n)}_{\mathrm{zz}}=\frac{1}{L}\int_0^L\epsilon_{\mathrm{zz}}(z-c_st)\cos(k_nz)dz$ accounting for the propagation of acoustic pulses through the ferromagnetic layer at the speed of sound $c_s$, where the dominant role of phase-matching conditions between magnons and phonons can be elucidated. In this paper we are going to study a special case of a ferromagnetic thin film and it comes out that in order to excite high-frequency exchange magnons of the order $n$, the Fourier spectra of acoustic pulses (in $k$-space, along the propagation direction $z$) should possess non-zero components at the respective wavevector $k_n$. As we are going to see in the next section, this condition is fulfilled for picosecond acoustic pulses generated by ultrashort laser pulses.

The system of equations (\ref{eq:dmxdt2},\ref{eq:dmydt2},\ref{eq:dmzdt2}) can be reduced to an equation of a harmonic oscillator for each magnon mode. Taking the time-derivative of Eq.~(\ref{eq:dmydt2}) and using Eqs.~(\ref{eq:dmxdt2}, \ref{eq:dmzdt2}) it is easy to obtain the following equation:   
\begin{equation}
\label{eq:oscillator}
\frac{d^{2}s^{(n)}_{\mathrm{y}}}{dt^{2}} + \omega^{2}_{n} s^{(n)}_{\mathrm{y}} =    \frac{2\gamma b_1\sin\theta\cos\theta}{M_0}\frac{d \epsilon^{(n)}_{\mathrm{zz}}(t)}{dt}\,,
\end{equation}
where $\omega_{n} = \omega(k_{n}) = \sqrt{-A_{12}^{\left( n \right)}A_{21}^{\left( n \right)}-A_{23}^{\left( n \right)}A_{32}^{\left( n \right)}}$ is the magnon frequency given by Eq.~\ref{eq:mFourier}. Taking into account that the most common experimental configuration for the polar magneto-optical Kerr effect measures the $z$-component of the magnetization vector, it is useful to derive an equation for $s^{(n)}_{\mathrm{z}}$. For that we extract $s^{(n)}_{\mathrm{y}}$ from Eq.~(\ref{eq:dmzdt2}) and substitute it in Eq.(~\ref{eq:oscillator}):
\begin{equation}
\label{eq:mz_from_oscil}
\frac{d^{3}s^{(n)}_{\mathrm{z}}}{dt^{3}} + \omega^{2}_{n} \frac{ds^{(n)}_{\mathrm{z}}}{dt} =    \frac{2 A_{32}^{\left( n \right)} \gamma b_1\sin\theta\cos\theta}{M_0}\frac{d \epsilon^{(n)}_{\mathrm{zz}}(t)}{dt}
\end{equation}
Integration over time leads to the usual oscillator equation:
\begin{equation}
\frac{d^{2}s^{(n)}_{\mathrm{z}}}{dt^{2}} + \omega^{2}_{n} s^{(n)}_{\mathrm{z}} =    \frac{2 A_{32}^{\left( n \right)} \gamma b_1\sin\theta\cos\theta}{M_0} \epsilon^{(n)}_{\mathrm{zz}}(t)
\end{equation}
By adding the damping term and the explicit expression for $A_{32}^{\left( n \right)}$ from Eq.~(27) we arrive at the most important analytical equation in this paper:   
\begin{equation}
\frac{d^{2}s^{(n)}_{\mathrm{z}}}{dt^{2}} + \alpha\omega_n\frac{ds^{(n)}_{\mathrm{z}}}{dt}+\omega^{2}_{n} s^{(n)}_{\mathrm{z}} =    \frac{ \mu_0 \gamma^2 b_1\sin (2\theta) (Dk^2_n\sin\theta+H\sin\xi)}{M_0 L}\int_0^L\epsilon_{\mathrm{zz}}(z,t)\cos(k_nz)dz\,.
\end{equation}
The magnon decay constant $\alpha\omega_n$, determined by the Gilbert damping parameter $\alpha$, is included in a phenomenological way based on the fact that the magnon lifetime $1/(\alpha\omega_n)$ is inversely proportional to its frequency, this scaling verified up to the THz frequencies \cite{razdolski2017nanoscale}. In spite of some possible deviations of this approximation for the in-plane geometry of the external magnetic field, this approximation appears to be accurate for high frequency exchange magnons under the tilted magnetic field \cite{Salikhov2019}. The time-dependent driving force on the right hand side displays a non-trivial dependence both on the tilt angles $\theta$ and $\xi$ as well as the magnon order $n$. Moreover, this equation as well as the entire theory are valid for arbitrary acoustic strains $\epsilon_{\mathrm{zz}}(z,t)$.

\section{Results of numerical simulations}
 Here we apply the developed theory to calculate the magnetization dynamics in a polycrystalline nickel film induced by picosecond acoustic pulses. Since the frequency interval between the neighbouring  magnon modes is mainly determined by the film thickness and their widths in the Fourier spectra are given by inverse lifetimes, here we consider a rather small film thickness of $L$=30~nm. It allows individual magnon modes to be resolved in frequency domain. Moreover, the velocity of longitudinal acoustic waves in polycrystalline nickel $c_s$=5.6~nm/ps provides a rather short acoustic travel time through the sample of $L/c_s$=5.4~ps, which is short enough compared to the characteristic magnon lifetimes. It makes it possible to observe the magnon dynamics after the acoustic pulse escaped from the ferromagnetic layer. The damping-free system of ordinary differential equations (\ref{eq:dmxdt2}-\ref{eq:dmzdt2}) is solved by the 4th-order Runge-Kutta method. At each Runge-Kutta integration time step $\Delta t$ all magnon components $s^{(n)}_i$ are multiplied by $\exp(-\alpha\omega_n\Delta t)$ providing magnon life times nearly identical to those obtained from the numerical solutions of the LLG equations (\ref{eq:LLG}). In order to calculate the magnetization dynamics $m_i(z,t)$, we are summing up over all magnon modes using Eq. (\ref{eq:Fourier}) truncated at N=20 as the amplitudes of higher magnon modes are negligibly small for all simulations discussed in this manuscript. For numerical calculations for ferromagnetic nickel we used nickel the Gilbert damping parameter $\alpha=0.05$ \cite{van2002all} and magnetoelastic coupling constant $b_1=10^7$~J/m$^3$ \cite{getzlaff2008}.

\begin{figure}[ht]
	\centering
	\includegraphics[width=0.95\columnwidth]{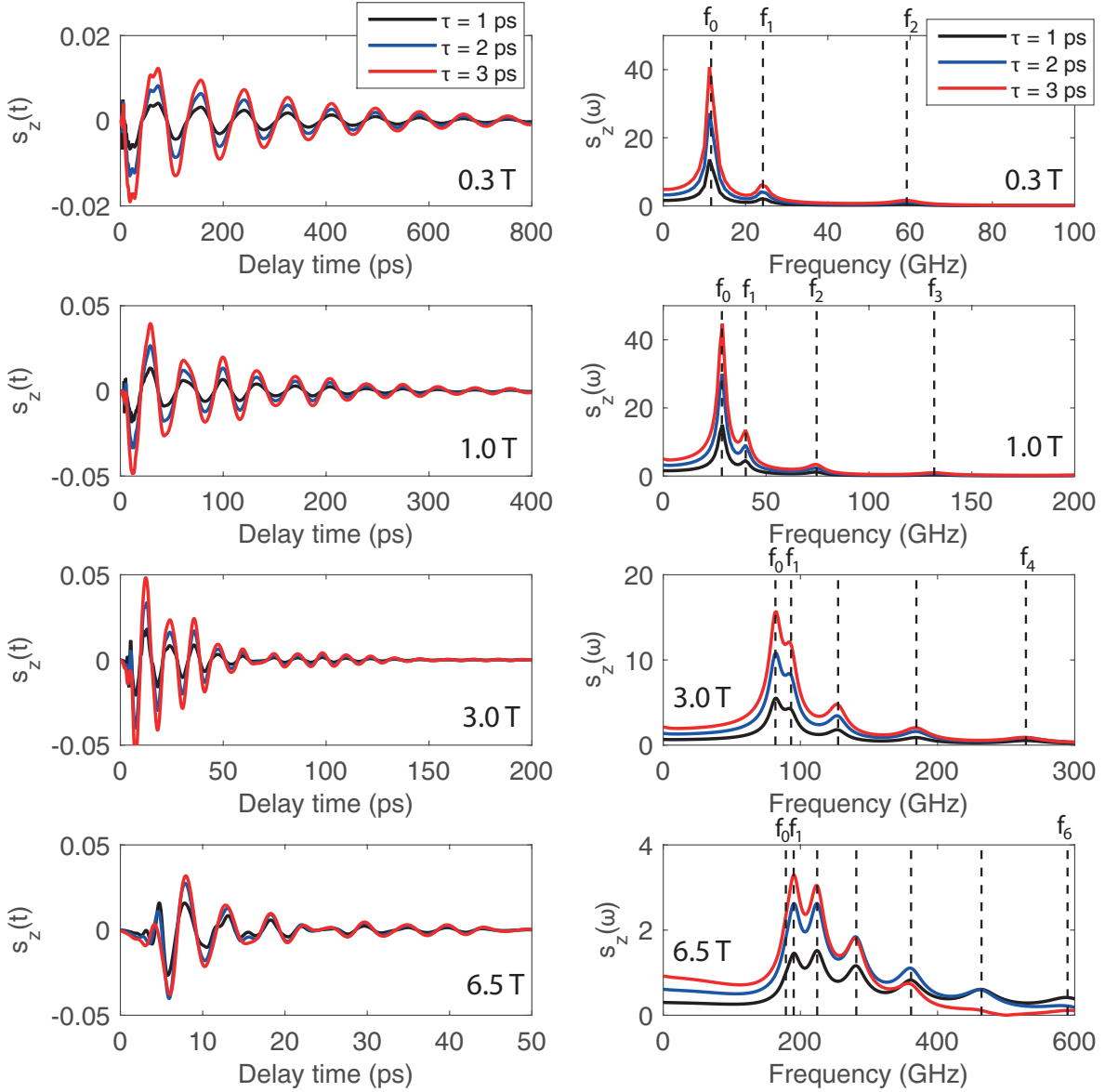}
	\caption{Variations of the magnetization component at the back interface $s_z(t,z=L)$ and their Fourier spectra obtained for three different acoustic pulse durations $\tau$=1,~2 and 3~ps and four values of an external magnetic field $\mu_{0}H$=0.3,~1,~3 and 6.5~T, respectively. The amplitude of acoustic pulses is $0.5\%$, the magnetic field is tilted by $\xi$=45$^{\circ}$.}
	\label{Fig2new}
\end{figure}

\begin{figure}[ht]
	\centering
	\includegraphics[width=0.95\columnwidth]{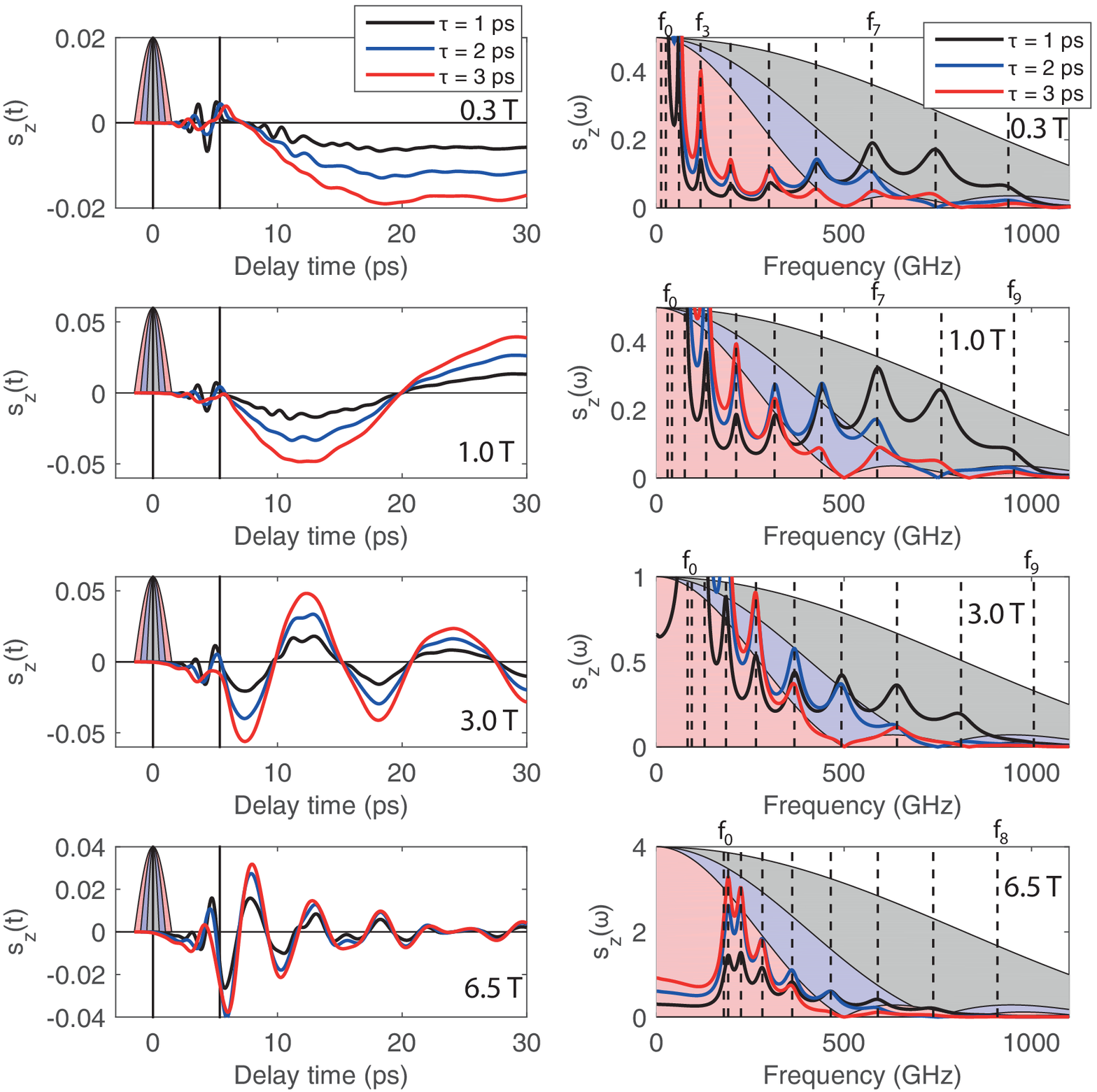}
	\caption{Variations of the magnetization component at the back interface $s_z(t,z=L)$ at their Fourier spectra for three different acoustic pulse durations $\tau$=1,~2 and 3~ps and four values of an external magnetic field $\mu_{0}H$=0.3,~1,~3 and 6.5~T, respectively. The shaded area represent the acoustic pulses and their spectra, respectively. Acoustic pulses injected at zero delay time escape from the nickel layer at 5.4~ps, as indicated by the dashed vertical lines. The amplitude of acoustic pulses is $0.5\%$, the magnetic field is tilted by $\xi$=45$^{\circ}$.}
	\label{Fig3new}
\end{figure}

Being aware of the importance of the acoustic bandwidth from the analysis of Eqs.~(\ref{eq:dmxdt2}-\ref{eq:dmzdt2}), we have performed the numerical simulations for three values of the acoustic pulse duration: 1,~2 and 3~ps corresponding to their spatial width in nickel of 5.6,~11.2 and 16.8~nm, respectively. Ultrashort acoustic pulses with 2-3~ps duration can be routinely generated in thin metal samples excited by femtosecond laser pulses \cite{ThomsenPRB86,MankeAPL2013,TemnovNatureComm2013}. Even shorter acoustic pulses with ultimate pulse durations in the deeply subpicosecond range have been observed in form of acoustic solitons resulting from the nonlinear propagation effects of ultrashort acoustic pulses in crystalline solids at cryogenic temperatures \cite{vanCapelUltrasonics2015}. Whereas subpicosecond soliton strains can get as high as $2\times 10^{-3}$, acoustic pulses generated in ferromagentic thin films can reach $10^{-2}=1\%$ amplitudes \cite{TemnovNatureComm2013}. As such, large amplitude ultrashort acoustic pulses can be routinely obtained in every laboratory for femtosecond laser spectroscopy.   

At present we do not discuss the details of acoustic injection in a ferromagnetic thin film and just assume it is sandwiched between two acoustically matched nonmagnetic materials allowing for an ultrashort acoustic pulse to be injected through the front interface ($z$=0) at zero time and leave it through the back interface ($z=L$) 5.4 picoseconds later. We have analyzed the magnetization dynamics at both interfaces for different amplitudes of the external magnetic field. In case of a thin sample, the magnon dynamics at both interfaces are quite similar. For this reason, in this paper we focus on the data obtained for the normal component of the magnetization $s_z(t)$ at the back interface and their Fourier spectra calculated for three acoustic pulse durations $\tau$=1,~2 and 3~ps (with pulse shapes given by a half of the period of the cosine function) and four values of an external magnetic field of 0.3,~1,~3, and 6.5~Tesla, respectively (see Fig.~\ref{Fig2new}). 
For small value of the magnetic field $\mu_0H=0.3$~T the Fourier spectra for all three pulse durations consist of a strong FMR peak at 11.6~GHz frequency and high-frequency magnon peaks. Being much weaker as compared to the FMR, the peaks for two lowest exchange modes at $f_1$=24.2~GHz and $f_2$=59.2~GHz are still visible in the Fourier spectra. As the magnetic field is increased, all frequencies are shifted up and the magnon amplitudes increase relatively to the FMR frequency. At the magnetic field of $\mu_0H=3.0$~T, the amplitude of he first magnon $f_1=93.2$~GHz almost reaches the strength of the FMR mode oscillating at $f_0=81.8$~GHz. This observation correlates very well with the concept in Fig.~1(b) showing that the magnon dispersion crosses the phonon dispersion precisely at the frequency of the first magnon mode, i.e. the phonon-magnon phase-matching condition is fulfilled.
For $\mu_0H=6.5$~T, the phonon and magnon dispersions cross at two points, which are close to the frequencies of the 3rd and the 5th magnon modes, and are almost parallel over a larger frequency range. The multitude of higher order magnons is excited in this case.
The mechanisms of magnon excitation and the acoustic pulse width can be understood, if one looks at the time dependencies and the Fourier spectra in Fig.~\ref{Fig3new}. 
At $\mu_0H=3.0$~T, the magnetization dynamics starts before at the delay time around 2~ps, i.e. well before the arrival time of an acoustic pulse at the back interface at 5.4~ps, marked by a black vertical line in all time dependencies in Fig.~\ref{Fig3new}. This initial part of magnetization dynamics oscillates at the characteristic time scale of approximately 1...2~ps. The corresponding Fourier spectrum on the right hand side (note the difference in the vertical scaling as compared to Fig.~\ref{Fig2new}) indeed contains several spectrally overlapping high frequency magnons peaking around the 7th magnon mode at $f_7=573$~GHz. This frequency range corresponds to the the second crossing point between phonon and magnon dispersions in Fig.~1(b) at small magnetic fields. These high frequency magnons are characterized by a higher group velocity. Therefore they arrive at the back interface earlier than the acoustic pulse that has generated them.    
This interpretation is consistent with the results for higher magnetic fields. Notably at $\mu_0H=6.5$~T the entire magnetization dynamics can be seen as a {\it chirped} magnon pulse consisting of the high frequency components arriving at short delay times and low frequency components arriving later.   

The role of the acoustic pulse duration becomes evident from an obvious visual correlation of the excited magnon spectra and the Fourier spectra of ultrashort acoustic pulses marked as colored shaded area on the right hand side of Fig.~3: the shorter the pulses the broader their spectrum, the more efficiently they excite high frequency magnons. As discussed earlier, this conclusion is in line with the results of our analytical considerations suggesting that the external driving force for magnons is proportional to the Fourier component of an acoustic pulse at the magnon frequency. At this point it is worth mentioning that the dependence of FMR amplitude on the acoustic pulse duration is opposite: longer acoustic pulses possess a larger {\it pulse area} \cite{KovalenkoPRL2013} resulting in a larger amplitude of an acoustically generated FMR precession. This behavior is evident in the Fourier spectra of Fig.~\ref{Fig2new}, both for the FMR and low-frequency exchange magnons.

\section{Summary and conclusions}
The main result of this paper is that ultrashort acoustic pulses propagating through thin ferromagnetic samples must excite not only the FMR precession but also high-order exchange magnons falling within the spectral bandwidth of acoustic pulses. Although the efficiency of magnon excitation is naturally enhanced when phonon-magnon phase-matching conditions are fulfilled, our simulations show that exchange magnons with measurable amplitudes get excited even when phonon and magnon dispersion do not cross. It is rather the acoustic bandwidth than phase matching that determines the excitation efficiency of exchange magnons. The optimum conditions for ultrafast magnetoelastic generation of exchange magnons can be elucidated from systematic numerical simulations as a function of multiple physical parameters such as the exchange stiffness, Gilbert damping, sample thickness, acoustic pulse duration etc.  Such analysis could be quite helpful in a view of the  experiments evidencing an ultrafast optical excitation of the exchange magnons in ferromagnetic thin films \cite{van2002all,Salikhov2019}, where possible contributions of magnetoelastic excitation of exchange magnons are masked by the dominant mechanism of ultrafast demagnetization \cite{beaurepaire1996ultrafast}. Applications of our simple theory to the experimental investigations on thick ferromagnetic films, characterized by a quasi-continuous magnon spectrum and suggesting the physical interpretation in terms of propagating magnon pulses, can be envisaged \cite{bombeck2012excitation,kim2017magnetization}. 

\bigskip
\section*{Acknowledgements}
Funding through Agence Nationale de la Recherche under grant ''PPMI-NANO'' (ANR-15-CE24-0032 and DFG SE2443/2), Strat\'{e}gie internationale NNN-Telecom de la R\'{e}gion Pays de La Loire, Alexander von Humboldt Stiftung, PRC CNRS-RFBR "Acousto-magneto-plasmonics" (grant number 17-57-150001) and Act 211 Government of the Russian Federation (contract 02.A03.21.0011) is greatfully acknowledged.

\section*{References}	
\bibliography{bib_ExchangeMagnons}

\begin{thebibliography}{41}
\expandafter\ifx\csname natexlab\endcsname\relax\def\natexlab#1{#1}\fi
\expandafter\ifx\csname bibnamefont\endcsname\relax
  \def\bibnamefont#1{#1}\fi
\expandafter\ifx\csname bibfnamefont\endcsname\relax
  \def\bibfnamefont#1{#1}\fi
\expandafter\ifx\csname citenamefont\endcsname\relax
  \def\citenamefont#1{#1}\fi
\expandafter\ifx\csname url\endcsname\relax
  \def\url#1{\texttt{#1}}\fi
\expandafter\ifx\csname urlprefix\endcsname\relax\def\urlprefix{URL }\fi
\providecommand{\bibinfo}[2]{#2}
\providecommand{\eprint}[2][]{\url{#2}}

\bibitem[{\citenamefont{Beaurepaire et~al.}(1996)\citenamefont{Beaurepaire,
  Merle, Daunois, and Bigot}}]{beaurepaire1996ultrafast}
\bibinfo{author}{\bibfnamefont{E.}~\bibnamefont{Beaurepaire}},
  \bibinfo{author}{\bibfnamefont{J.-C.} \bibnamefont{Merle}},
  \bibinfo{author}{\bibfnamefont{A.}~\bibnamefont{Daunois}}, \bibnamefont{and}
  \bibinfo{author}{\bibfnamefont{J.-Y.} \bibnamefont{Bigot}},
  \bibinfo{journal}{Phys. Rev. Lett.} \textbf{\bibinfo{volume}{76}},
  \bibinfo{pages}{4250} (\bibinfo{year}{1996}),
  \urlprefix\url{https://link.aps.org/doi/10.1103/PhysRevLett.76.4250}.

\bibitem[{\citenamefont{Koopmans et~al.}(2000)\citenamefont{Koopmans,
  Van~Kampen, Kohlhepp, and De~Jonge}}]{koopmans2000ultrafast}
\bibinfo{author}{\bibfnamefont{B.}~\bibnamefont{Koopmans}},
  \bibinfo{author}{\bibfnamefont{M.}~\bibnamefont{Van~Kampen}},
  \bibinfo{author}{\bibfnamefont{J.}~\bibnamefont{Kohlhepp}}, \bibnamefont{and}
  \bibinfo{author}{\bibfnamefont{W.}~\bibnamefont{De~Jonge}},
  \bibinfo{journal}{Phys. Rev. Lett.} \textbf{\bibinfo{volume}{85}},
  \bibinfo{pages}{844} (\bibinfo{year}{2000}),
  \urlprefix\url{https://link.aps.org/doi/10.1103/PhysRevLett.85.844}.

\bibitem[{\citenamefont{Van~Kampen et~al.}(2002)\citenamefont{Van~Kampen,
  Jozsa, Kohlhepp, LeClair, Lagae, De~Jonge, and Koopmans}}]{van2002all}
\bibinfo{author}{\bibfnamefont{M.}~\bibnamefont{Van~Kampen}},
  \bibinfo{author}{\bibfnamefont{C.}~\bibnamefont{Jozsa}},
  \bibinfo{author}{\bibfnamefont{J.}~\bibnamefont{Kohlhepp}},
  \bibinfo{author}{\bibfnamefont{P.}~\bibnamefont{LeClair}},
  \bibinfo{author}{\bibfnamefont{L.}~\bibnamefont{Lagae}},
  \bibinfo{author}{\bibfnamefont{W.}~\bibnamefont{De~Jonge}}, \bibnamefont{and}
  \bibinfo{author}{\bibfnamefont{B.}~\bibnamefont{Koopmans}},
  \bibinfo{journal}{Phys. Rev. Lett.} \textbf{\bibinfo{volume}{88}},
  \bibinfo{pages}{227201} (\bibinfo{year}{2002}),
  \urlprefix\url{https://link.aps.org/doi/10.1103/PhysRevLett.88.227201}.

\bibitem[{\citenamefont{Guidoni et~al.}(2002)\citenamefont{Guidoni,
  Beaurepaire, and Bigot}}]{guidoni2002magneto}
\bibinfo{author}{\bibfnamefont{L.}~\bibnamefont{Guidoni}},
  \bibinfo{author}{\bibfnamefont{E.}~\bibnamefont{Beaurepaire}},
  \bibnamefont{and} \bibinfo{author}{\bibfnamefont{J.-Y.} \bibnamefont{Bigot}},
  \bibinfo{journal}{Phys. Rev. Lett.} \textbf{\bibinfo{volume}{89}},
  \bibinfo{pages}{017401} (\bibinfo{year}{2002}),
  \urlprefix\url{https://link.aps.org/doi/10.1103/PhysRevLett.89.017401}.

\bibitem[{\citenamefont{Zhang et~al.}(2002)\citenamefont{Zhang, Nurmikko,
  Anguelouch, Xiao, and Gupta}}]{zhang2002coherent}
\bibinfo{author}{\bibfnamefont{Q.}~\bibnamefont{Zhang}},
  \bibinfo{author}{\bibfnamefont{A.~V.} \bibnamefont{Nurmikko}},
  \bibinfo{author}{\bibfnamefont{A.}~\bibnamefont{Anguelouch}},
  \bibinfo{author}{\bibfnamefont{G.}~\bibnamefont{Xiao}}, \bibnamefont{and}
  \bibinfo{author}{\bibfnamefont{A.}~\bibnamefont{Gupta}},
  \bibinfo{journal}{Phys. Rev. Lett.} \textbf{\bibinfo{volume}{89}},
  \bibinfo{pages}{177402} (\bibinfo{year}{2002}),
  \urlprefix\url{https://link.aps.org/doi/10.1103/PhysRevLett.89.177402}.

\bibitem[{\citenamefont{Vomir et~al.}(2005)\citenamefont{Vomir, Andrade,
  Guidoni, Beaurepaire, and Bigot}}]{vomir2005real}
\bibinfo{author}{\bibfnamefont{M.}~\bibnamefont{Vomir}},
  \bibinfo{author}{\bibfnamefont{L.}~\bibnamefont{Andrade}},
  \bibinfo{author}{\bibfnamefont{L.}~\bibnamefont{Guidoni}},
  \bibinfo{author}{\bibfnamefont{E.}~\bibnamefont{Beaurepaire}},
  \bibnamefont{and} \bibinfo{author}{\bibfnamefont{J.-Y.} \bibnamefont{Bigot}},
  \bibinfo{journal}{Phys. Rev. Lett.} \textbf{\bibinfo{volume}{94}},
  \bibinfo{pages}{237601} (\bibinfo{year}{2005}),
  \urlprefix\url{https://link.aps.org/doi/10.1103/PhysRevLett.94.237601}.

\bibitem[{\citenamefont{Bigot et~al.}(2005)\citenamefont{Bigot, Vomir, Andrade,
  and Beaurepaire}}]{bigot2005ultrafast}
\bibinfo{author}{\bibfnamefont{J.-Y.} \bibnamefont{Bigot}},
  \bibinfo{author}{\bibfnamefont{M.}~\bibnamefont{Vomir}},
  \bibinfo{author}{\bibfnamefont{L.}~\bibnamefont{Andrade}}, \bibnamefont{and}
  \bibinfo{author}{\bibfnamefont{E.}~\bibnamefont{Beaurepaire}},
  \bibinfo{journal}{Chemical physics} \textbf{\bibinfo{volume}{318}},
  \bibinfo{pages}{137} (\bibinfo{year}{2005}),
  \urlprefix\url{http://www.sciencedirect.com/science/article/pii/S030101040500248X}.

\bibitem[{\citenamefont{Kimel et~al.}(2005)\citenamefont{Kimel, Kirilyuk,
  Usachev, Pisarev et~al.}}]{kimel2005ultrafast}
\bibinfo{author}{\bibfnamefont{A.}~\bibnamefont{Kimel}},
  \bibinfo{author}{\bibfnamefont{A.}~\bibnamefont{Kirilyuk}},
  \bibinfo{author}{\bibfnamefont{P.}~\bibnamefont{Usachev}},
  \bibinfo{author}{\bibfnamefont{R.}~\bibnamefont{Pisarev}},
  \bibnamefont{et~al.}, \bibinfo{journal}{Nature}
  \textbf{\bibinfo{volume}{435}}, \bibinfo{pages}{655} (\bibinfo{year}{2005}),
  \urlprefix\url{https://doi.org/10.1038/nature03564}.

\bibitem[{\citenamefont{Malinowski et~al.}(2008)\citenamefont{Malinowski,
  Dalla~Longa, Rietjens, Paluskar, Huijink, Swagten, and
  Koopmans}}]{malinowski2008control}
\bibinfo{author}{\bibfnamefont{G.}~\bibnamefont{Malinowski}},
  \bibinfo{author}{\bibfnamefont{F.}~\bibnamefont{Dalla~Longa}},
  \bibinfo{author}{\bibfnamefont{J.}~\bibnamefont{Rietjens}},
  \bibinfo{author}{\bibfnamefont{P.}~\bibnamefont{Paluskar}},
  \bibinfo{author}{\bibfnamefont{R.}~\bibnamefont{Huijink}},
  \bibinfo{author}{\bibfnamefont{H.}~\bibnamefont{Swagten}}, \bibnamefont{and}
  \bibinfo{author}{\bibfnamefont{B.}~\bibnamefont{Koopmans}},
  \bibinfo{journal}{Nature Physics} \textbf{\bibinfo{volume}{4}},
  \bibinfo{pages}{855} (\bibinfo{year}{2008}),
  \urlprefix\url{https://doi.org/10.1038/nphys1092}.

\bibitem[{\citenamefont{Bigot et~al.}(2009)\citenamefont{Bigot, Vomir, and
  Beaurepaire}}]{bigot2009coherent}
\bibinfo{author}{\bibfnamefont{J.-Y.} \bibnamefont{Bigot}},
  \bibinfo{author}{\bibfnamefont{M.}~\bibnamefont{Vomir}}, \bibnamefont{and}
  \bibinfo{author}{\bibfnamefont{E.}~\bibnamefont{Beaurepaire}},
  \bibinfo{journal}{Nature Physics} \textbf{\bibinfo{volume}{5}},
  \bibinfo{pages}{515} (\bibinfo{year}{2009}),
  \urlprefix\url{https://doi.org/10.1038/nphys1285}.

\bibitem[{\citenamefont{Bovensiepen}(2009)}]{bovensiepen2009femtomagnetism}
\bibinfo{author}{\bibfnamefont{U.}~\bibnamefont{Bovensiepen}},
  \bibinfo{journal}{Nature Physics} \textbf{\bibinfo{volume}{5}},
  \bibinfo{pages}{461} (\bibinfo{year}{2009}),
  \urlprefix\url{https://doi.org/10.1038/nphys1322}.

\bibitem[{\citenamefont{Radu et~al.}(2009)\citenamefont{Radu, Woltersdorf,
  Kiessling, Melnikov, Bovensiepen, Thiele, and Back}}]{radu2009laser}
\bibinfo{author}{\bibfnamefont{I.}~\bibnamefont{Radu}},
  \bibinfo{author}{\bibfnamefont{G.}~\bibnamefont{Woltersdorf}},
  \bibinfo{author}{\bibfnamefont{M.}~\bibnamefont{Kiessling}},
  \bibinfo{author}{\bibfnamefont{A.}~\bibnamefont{Melnikov}},
  \bibinfo{author}{\bibfnamefont{U.}~\bibnamefont{Bovensiepen}},
  \bibinfo{author}{\bibfnamefont{J.-U.} \bibnamefont{Thiele}},
  \bibnamefont{and} \bibinfo{author}{\bibfnamefont{C.~H.} \bibnamefont{Back}},
  \bibinfo{journal}{Phys. Rev. Lett.} \textbf{\bibinfo{volume}{102}},
  \bibinfo{pages}{117201} (\bibinfo{year}{2009}),
  \urlprefix\url{https://link.aps.org/doi/10.1103/PhysRevLett.102.117201}.

\bibitem[{\citenamefont{Carpene et~al.}(2010)\citenamefont{Carpene, Mancini,
  Dazzi, Dallera, Puppin, and De~Silvestri}}]{carpene2010ultrafast}
\bibinfo{author}{\bibfnamefont{E.}~\bibnamefont{Carpene}},
  \bibinfo{author}{\bibfnamefont{E.}~\bibnamefont{Mancini}},
  \bibinfo{author}{\bibfnamefont{D.}~\bibnamefont{Dazzi}},
  \bibinfo{author}{\bibfnamefont{C.}~\bibnamefont{Dallera}},
  \bibinfo{author}{\bibfnamefont{E.}~\bibnamefont{Puppin}}, \bibnamefont{and}
  \bibinfo{author}{\bibfnamefont{S.}~\bibnamefont{De~Silvestri}},
  \bibinfo{journal}{Phys. Rev. B} \textbf{\bibinfo{volume}{81}},
  \bibinfo{pages}{060415} (\bibinfo{year}{2010}),
  \urlprefix\url{https://link.aps.org/doi/10.1103/PhysRevB.81.060415}.

\bibitem[{\citenamefont{Boeglin et~al.}(2010)\citenamefont{Boeglin,
  Beaurepaire, Halt{\'e}, L{\'o}pez-Flores, Stamm, Pontius, D{\"u}rr, and
  Bigot}}]{boeglin2010distinguishing}
\bibinfo{author}{\bibfnamefont{C.}~\bibnamefont{Boeglin}},
  \bibinfo{author}{\bibfnamefont{E.}~\bibnamefont{Beaurepaire}},
  \bibinfo{author}{\bibfnamefont{V.}~\bibnamefont{Halt{\'e}}},
  \bibinfo{author}{\bibfnamefont{V.}~\bibnamefont{L{\'o}pez-Flores}},
  \bibinfo{author}{\bibfnamefont{C.}~\bibnamefont{Stamm}},
  \bibinfo{author}{\bibfnamefont{N.}~\bibnamefont{Pontius}},
  \bibinfo{author}{\bibfnamefont{H.}~\bibnamefont{D{\"u}rr}}, \bibnamefont{and}
  \bibinfo{author}{\bibfnamefont{J.}~\bibnamefont{Bigot}},
  \bibinfo{journal}{Nature} \textbf{\bibinfo{volume}{465}},
  \bibinfo{pages}{458} (\bibinfo{year}{2010}),
  \urlprefix\url{https://doi.org/10.1038/nature09070}.

\bibitem[{\citenamefont{Scherbakov et~al.}(2010)\citenamefont{Scherbakov,
  Salasyuk, Akimov, Liu, Bombeck, Br\"uggemann, Yakovlev, Sapega, Furdyna, and
  Bayer}}]{Scherbakov2010}
\bibinfo{author}{\bibfnamefont{A.~V.} \bibnamefont{Scherbakov}},
  \bibinfo{author}{\bibfnamefont{A.~S.} \bibnamefont{Salasyuk}},
  \bibinfo{author}{\bibfnamefont{A.~V.} \bibnamefont{Akimov}},
  \bibinfo{author}{\bibfnamefont{X.}~\bibnamefont{Liu}},
  \bibinfo{author}{\bibfnamefont{M.}~\bibnamefont{Bombeck}},
  \bibinfo{author}{\bibfnamefont{C.}~\bibnamefont{Br\"uggemann}},
  \bibinfo{author}{\bibfnamefont{D.~R.} \bibnamefont{Yakovlev}},
  \bibinfo{author}{\bibfnamefont{V.~F.} \bibnamefont{Sapega}},
  \bibinfo{author}{\bibfnamefont{J.~K.} \bibnamefont{Furdyna}},
  \bibnamefont{and} \bibinfo{author}{\bibfnamefont{M.}~\bibnamefont{Bayer}},
  \bibinfo{journal}{Phys. Rev. Lett.} \textbf{\bibinfo{volume}{105}},
  \bibinfo{pages}{117204} (\bibinfo{year}{2010}),
  \urlprefix\url{https://link.aps.org/doi/10.1103/PhysRevLett.105.117204}.

\bibitem[{\citenamefont{Radu et~al.}(2011)\citenamefont{Radu, Vahaplar, Stamm,
  Kachel, Pontius, D{\"u}rr, Ostler, Barker, Evans, Chantrell
  et~al.}}]{radu2011transient}
\bibinfo{author}{\bibfnamefont{I.}~\bibnamefont{Radu}},
  \bibinfo{author}{\bibfnamefont{K.}~\bibnamefont{Vahaplar}},
  \bibinfo{author}{\bibfnamefont{C.}~\bibnamefont{Stamm}},
  \bibinfo{author}{\bibfnamefont{T.}~\bibnamefont{Kachel}},
  \bibinfo{author}{\bibfnamefont{N.}~\bibnamefont{Pontius}},
  \bibinfo{author}{\bibfnamefont{H.}~\bibnamefont{D{\"u}rr}},
  \bibinfo{author}{\bibfnamefont{T.}~\bibnamefont{Ostler}},
  \bibinfo{author}{\bibfnamefont{J.}~\bibnamefont{Barker}},
  \bibinfo{author}{\bibfnamefont{R.}~\bibnamefont{Evans}},
  \bibinfo{author}{\bibfnamefont{R.}~\bibnamefont{Chantrell}},
  \bibnamefont{et~al.}, \bibinfo{journal}{Nature}
  \textbf{\bibinfo{volume}{472}}, \bibinfo{pages}{205} (\bibinfo{year}{2011}),
  \urlprefix\url{https://doi.org/10.1038/nature09901}.

\bibitem[{\citenamefont{Rudolf et~al.}(2012)\citenamefont{Rudolf, Chan,
  Battiato, Adam, Shaw, Turgut, Maldonado, Mathias, Grychtol, Nembach
  et~al.}}]{rudolf2012ultrafast}
\bibinfo{author}{\bibfnamefont{D.}~\bibnamefont{Rudolf}},
  \bibinfo{author}{\bibfnamefont{L.-O.} \bibnamefont{Chan}},
  \bibinfo{author}{\bibfnamefont{M.}~\bibnamefont{Battiato}},
  \bibinfo{author}{\bibfnamefont{R.}~\bibnamefont{Adam}},
  \bibinfo{author}{\bibfnamefont{J.~M.} \bibnamefont{Shaw}},
  \bibinfo{author}{\bibfnamefont{E.}~\bibnamefont{Turgut}},
  \bibinfo{author}{\bibfnamefont{P.}~\bibnamefont{Maldonado}},
  \bibinfo{author}{\bibfnamefont{S.}~\bibnamefont{Mathias}},
  \bibinfo{author}{\bibfnamefont{P.}~\bibnamefont{Grychtol}},
  \bibinfo{author}{\bibfnamefont{H.~T.} \bibnamefont{Nembach}},
  \bibnamefont{et~al.}, \bibinfo{journal}{Nature communications}
  \textbf{\bibinfo{volume}{3}}, \bibinfo{pages}{1037} (\bibinfo{year}{2012}),
  \urlprefix\url{https://doi.org/10.1038/ncomms2029}.

\bibitem[{\citenamefont{Bombeck et~al.}(2013)\citenamefont{Bombeck, J\"ager,
  Scherbakov, Linnik, Yakovlev, Liu, Furdyna, Akimov, and
  Bayer}}]{bombeck2013magnetization}
\bibinfo{author}{\bibfnamefont{M.}~\bibnamefont{Bombeck}},
  \bibinfo{author}{\bibfnamefont{J.~V.} \bibnamefont{J\"ager}},
  \bibinfo{author}{\bibfnamefont{A.~V.} \bibnamefont{Scherbakov}},
  \bibinfo{author}{\bibfnamefont{T.}~\bibnamefont{Linnik}},
  \bibinfo{author}{\bibfnamefont{D.~R.} \bibnamefont{Yakovlev}},
  \bibinfo{author}{\bibfnamefont{X.}~\bibnamefont{Liu}},
  \bibinfo{author}{\bibfnamefont{J.~K.} \bibnamefont{Furdyna}},
  \bibinfo{author}{\bibfnamefont{A.~V.} \bibnamefont{Akimov}},
  \bibnamefont{and} \bibinfo{author}{\bibfnamefont{M.}~\bibnamefont{Bayer}},
  \bibinfo{journal}{Phys. Rev. B} \textbf{\bibinfo{volume}{87}},
  \bibinfo{pages}{060302} (\bibinfo{year}{2013}),
  \urlprefix\url{https://link.aps.org/doi/10.1103/PhysRevB.87.060302}.

\bibitem[{\citenamefont{Kim et~al.}(2012)\citenamefont{Kim, Vomir, and
  Bigot}}]{kim2012ultrafast}
\bibinfo{author}{\bibfnamefont{J.-W.} \bibnamefont{Kim}},
  \bibinfo{author}{\bibfnamefont{M.}~\bibnamefont{Vomir}}, \bibnamefont{and}
  \bibinfo{author}{\bibfnamefont{J.-Y.} \bibnamefont{Bigot}},
  \bibinfo{journal}{Phys. Rev. Lett.} \textbf{\bibinfo{volume}{109}},
  \bibinfo{pages}{166601} (\bibinfo{year}{2012}),
  \urlprefix\url{https://link.aps.org/doi/10.1103/PhysRevLett.109.166601}.

\bibitem[{\citenamefont{Kim et~al.}(2015)\citenamefont{Kim, Vomir, and
  Bigot}}]{kim2015controlling}
\bibinfo{author}{\bibfnamefont{J.-W.} \bibnamefont{Kim}},
  \bibinfo{author}{\bibfnamefont{M.}~\bibnamefont{Vomir}}, \bibnamefont{and}
  \bibinfo{author}{\bibfnamefont{J.-Y.} \bibnamefont{Bigot}},
  \bibinfo{journal}{Scientific reports} \textbf{\bibinfo{volume}{5}},
  \bibinfo{pages}{8511} (\bibinfo{year}{2015}),
  \urlprefix\url{https://doi.org/10.1038/srep08511}.

\bibitem[{\citenamefont{Kim and Bigot}(2017)}]{kim2017magnetization}
\bibinfo{author}{\bibfnamefont{J.-W.} \bibnamefont{Kim}} \bibnamefont{and}
  \bibinfo{author}{\bibfnamefont{J.-Y.} \bibnamefont{Bigot}},
  \bibinfo{journal}{Phys. Rev. B} \textbf{\bibinfo{volume}{95}},
  \bibinfo{pages}{144422} (\bibinfo{year}{2017}),
  \urlprefix\url{https://link.aps.org/doi/10.1103/PhysRevB.95.144422}.

\bibitem[{\citenamefont{Kirilyuk et~al.}(2010)\citenamefont{Kirilyuk, Kimel,
  and Rasing}}]{kirilyuk2010ultrafast}
\bibinfo{author}{\bibfnamefont{A.}~\bibnamefont{Kirilyuk}},
  \bibinfo{author}{\bibfnamefont{A.~V.} \bibnamefont{Kimel}}, \bibnamefont{and}
  \bibinfo{author}{\bibfnamefont{T.}~\bibnamefont{Rasing}},
  \bibinfo{journal}{Rev. Mod. Phys.} \textbf{\bibinfo{volume}{82}},
  \bibinfo{pages}{2731} (\bibinfo{year}{2010}),
  \urlprefix\url{https://link.aps.org/doi/10.1103/RevModPhys.82.2731}.

\bibitem[{\citenamefont{Thevenard et~al.}(2010)\citenamefont{Thevenard,
  Peronne, Gourdon, Testelin, Cubukcu, Charron, Vincent, Lema\^{\i}tre, and
  Perrin}}]{Thevenard2010}
\bibinfo{author}{\bibfnamefont{L.}~\bibnamefont{Thevenard}},
  \bibinfo{author}{\bibfnamefont{E.}~\bibnamefont{Peronne}},
  \bibinfo{author}{\bibfnamefont{C.}~\bibnamefont{Gourdon}},
  \bibinfo{author}{\bibfnamefont{C.}~\bibnamefont{Testelin}},
  \bibinfo{author}{\bibfnamefont{M.}~\bibnamefont{Cubukcu}},
  \bibinfo{author}{\bibfnamefont{E.}~\bibnamefont{Charron}},
  \bibinfo{author}{\bibfnamefont{S.}~\bibnamefont{Vincent}},
  \bibinfo{author}{\bibfnamefont{A.}~\bibnamefont{Lema\^{\i}tre}},
  \bibnamefont{and} \bibinfo{author}{\bibfnamefont{B.}~\bibnamefont{Perrin}},
  \bibinfo{journal}{Phys. Rev. B} \textbf{\bibinfo{volume}{82}},
  \bibinfo{pages}{104422} (\bibinfo{year}{2010}),
  \urlprefix\url{https://link.aps.org/doi/10.1103/PhysRevB.82.104422}.

\bibitem[{\citenamefont{Bombeck et~al.}(2012)\citenamefont{Bombeck, Salasyuk,
  Glavin, Scherbakov, Br{\"u}ggemann, Yakovlev, Sapega, Liu, Furdyna, Akimov
  et~al.}}]{bombeck2012excitation}
\bibinfo{author}{\bibfnamefont{M.}~\bibnamefont{Bombeck}},
  \bibinfo{author}{\bibfnamefont{A.}~\bibnamefont{Salasyuk}},
  \bibinfo{author}{\bibfnamefont{B.}~\bibnamefont{Glavin}},
  \bibinfo{author}{\bibfnamefont{A.}~\bibnamefont{Scherbakov}},
  \bibinfo{author}{\bibfnamefont{C.}~\bibnamefont{Br{\"u}ggemann}},
  \bibinfo{author}{\bibfnamefont{D.}~\bibnamefont{Yakovlev}},
  \bibinfo{author}{\bibfnamefont{V.}~\bibnamefont{Sapega}},
  \bibinfo{author}{\bibfnamefont{X.}~\bibnamefont{Liu}},
  \bibinfo{author}{\bibfnamefont{J.}~\bibnamefont{Furdyna}},
  \bibinfo{author}{\bibfnamefont{A.}~\bibnamefont{Akimov}},
  \bibnamefont{et~al.}, \bibinfo{journal}{Phys. Rev. B}
  \textbf{\bibinfo{volume}{85}}, \bibinfo{pages}{195324}
  (\bibinfo{year}{2012}),
  \urlprefix\url{https://link.aps.org/doi/10.1103/PhysRevB.85.195324}.

\bibitem[{\citenamefont{Temnov}(2012)}]{temnov2012ultrafast}
\bibinfo{author}{\bibfnamefont{V.~V.} \bibnamefont{Temnov}},
  \bibinfo{journal}{Nature Photonics} \textbf{\bibinfo{volume}{6}},
  \bibinfo{pages}{728} (\bibinfo{year}{2012}),
  \urlprefix\url{https://doi.org/10.1038/nphoton.2012.220}.

\bibitem[{\citenamefont{Janu{\v{s}}onis
  et~al.}(2016)\citenamefont{Janu{\v{s}}onis, Chang, Jansma, Gatilova, Vlasov,
  Lomonosov, Temnov, and Tobey}}]{januvsonis2016ultrafast}
\bibinfo{author}{\bibfnamefont{J.}~\bibnamefont{Janu{\v{s}}onis}},
  \bibinfo{author}{\bibfnamefont{C.-L.} \bibnamefont{Chang}},
  \bibinfo{author}{\bibfnamefont{T.}~\bibnamefont{Jansma}},
  \bibinfo{author}{\bibfnamefont{A.}~\bibnamefont{Gatilova}},
  \bibinfo{author}{\bibfnamefont{V.}~\bibnamefont{Vlasov}},
  \bibinfo{author}{\bibfnamefont{A.}~\bibnamefont{Lomonosov}},
  \bibinfo{author}{\bibfnamefont{V.}~\bibnamefont{Temnov}}, \bibnamefont{and}
  \bibinfo{author}{\bibfnamefont{R.}~\bibnamefont{Tobey}},
  \bibinfo{journal}{Phys. Rev. B} \textbf{\bibinfo{volume}{94}},
  \bibinfo{pages}{024415} (\bibinfo{year}{2016}),
  \urlprefix\url{https://link.aps.org/doi/10.1103/PhysRevB.94.024415}.

\bibitem[{\citenamefont{Chang et~al.}(2017)\citenamefont{Chang, Lomonosov,
  Janusonis, Vlasov, Temnov, and Tobey}}]{chang2017parametric}
\bibinfo{author}{\bibfnamefont{C.}~\bibnamefont{Chang}},
  \bibinfo{author}{\bibfnamefont{A.}~\bibnamefont{Lomonosov}},
  \bibinfo{author}{\bibfnamefont{J.}~\bibnamefont{Janusonis}},
  \bibinfo{author}{\bibfnamefont{V.}~\bibnamefont{Vlasov}},
  \bibinfo{author}{\bibfnamefont{V.}~\bibnamefont{Temnov}}, \bibnamefont{and}
  \bibinfo{author}{\bibfnamefont{R.}~\bibnamefont{Tobey}},
  \bibinfo{journal}{Phys. Rev. B} \textbf{\bibinfo{volume}{95}},
  \bibinfo{pages}{060409} (\bibinfo{year}{2017}),
  \urlprefix\url{https://link.aps.org/doi/10.1103/PhysRevB.95.060409}.

\bibitem[{\citenamefont{N{\v{e}}mec et~al.}(2012)\citenamefont{N{\v{e}}mec,
  Rozkotov{\'a}, Tesa{\v{r}}ov{\'a}, Troj{\'a}nek, De~Ranieri, Olejn{\'\i}k,
  Zemen, Nov{\'a}k, Cukr, Mal{\'y} et~al.}}]{nemec2012experimental}
\bibinfo{author}{\bibfnamefont{P.}~\bibnamefont{N{\v{e}}mec}},
  \bibinfo{author}{\bibfnamefont{E.}~\bibnamefont{Rozkotov{\'a}}},
  \bibinfo{author}{\bibfnamefont{N.}~\bibnamefont{Tesa{\v{r}}ov{\'a}}},
  \bibinfo{author}{\bibfnamefont{F.}~\bibnamefont{Troj{\'a}nek}},
  \bibinfo{author}{\bibfnamefont{E.}~\bibnamefont{De~Ranieri}},
  \bibinfo{author}{\bibfnamefont{K.}~\bibnamefont{Olejn{\'\i}k}},
  \bibinfo{author}{\bibfnamefont{J.}~\bibnamefont{Zemen}},
  \bibinfo{author}{\bibfnamefont{V.}~\bibnamefont{Nov{\'a}k}},
  \bibinfo{author}{\bibfnamefont{M.}~\bibnamefont{Cukr}},
  \bibinfo{author}{\bibfnamefont{P.}~\bibnamefont{Mal{\'y}}},
  \bibnamefont{et~al.}, \bibinfo{journal}{Nature physics}
  \textbf{\bibinfo{volume}{8}}, \bibinfo{pages}{411} (\bibinfo{year}{2012}),
  \urlprefix\url{https://doi.org/10.1038/nphys2279}.

\bibitem[{\citenamefont{Schellekens et~al.}(2014)\citenamefont{Schellekens,
  Kuiper, de~Wit, and Koopmans}}]{Schellekens2014}
\bibinfo{author}{\bibfnamefont{A.~J.} \bibnamefont{Schellekens}},
  \bibinfo{author}{\bibfnamefont{K.~C.} \bibnamefont{Kuiper}},
  \bibinfo{author}{\bibfnamefont{R.}~\bibnamefont{de~Wit}}, \bibnamefont{and}
  \bibinfo{author}{\bibfnamefont{B.}~\bibnamefont{Koopmans}},
  \bibinfo{journal}{Nature Communications} \textbf{\bibinfo{volume}{5}},
  \bibinfo{pages}{4333} (\bibinfo{year}{2014}),
  \urlprefix\url{https://doi.org/10.1038/ncomms5333}.

\bibitem[{\citenamefont{Razdolski et~al.}(2017)\citenamefont{Razdolski,
  Alekhin, Ilin, Meyburg, Roddatis, Diesing, Bovensiepen, and
  Melnikov}}]{razdolski2017nanoscale}
\bibinfo{author}{\bibfnamefont{I.}~\bibnamefont{Razdolski}},
  \bibinfo{author}{\bibfnamefont{A.}~\bibnamefont{Alekhin}},
  \bibinfo{author}{\bibfnamefont{N.}~\bibnamefont{Ilin}},
  \bibinfo{author}{\bibfnamefont{J.~P.} \bibnamefont{Meyburg}},
  \bibinfo{author}{\bibfnamefont{V.}~\bibnamefont{Roddatis}},
  \bibinfo{author}{\bibfnamefont{D.}~\bibnamefont{Diesing}},
  \bibinfo{author}{\bibfnamefont{U.}~\bibnamefont{Bovensiepen}},
  \bibnamefont{and} \bibinfo{author}{\bibfnamefont{A.}~\bibnamefont{Melnikov}},
  \bibinfo{journal}{Nature communications} \textbf{\bibinfo{volume}{8}},
  \bibinfo{pages}{15007} (\bibinfo{year}{2017}),
  \urlprefix\url{https://doi.org/10.1038/ncomms15007}.

\bibitem[{\citenamefont{Alekhin et~al.}(2017)\citenamefont{Alekhin, Razdolski,
  Ilin, Meyburg, Diesing, Roddatis, Rungger, Stamenova, Sanvito, Bovensiepen
  et~al.}}]{alekhin2017femtosecond}
\bibinfo{author}{\bibfnamefont{A.}~\bibnamefont{Alekhin}},
  \bibinfo{author}{\bibfnamefont{I.}~\bibnamefont{Razdolski}},
  \bibinfo{author}{\bibfnamefont{N.}~\bibnamefont{Ilin}},
  \bibinfo{author}{\bibfnamefont{J.~P.} \bibnamefont{Meyburg}},
  \bibinfo{author}{\bibfnamefont{D.}~\bibnamefont{Diesing}},
  \bibinfo{author}{\bibfnamefont{V.}~\bibnamefont{Roddatis}},
  \bibinfo{author}{\bibfnamefont{I.}~\bibnamefont{Rungger}},
  \bibinfo{author}{\bibfnamefont{M.}~\bibnamefont{Stamenova}},
  \bibinfo{author}{\bibfnamefont{S.}~\bibnamefont{Sanvito}},
  \bibinfo{author}{\bibfnamefont{U.}~\bibnamefont{Bovensiepen}},
  \bibnamefont{et~al.}, \bibinfo{journal}{Phys. Rev. Lett.}
  \textbf{\bibinfo{volume}{119}}, \bibinfo{pages}{017202}
  (\bibinfo{year}{2017}),
  \urlprefix\url{https://link.aps.org/doi/10.1103/PhysRevLett.119.017202}.

\bibitem[{\citenamefont{Alekhin et~al.}(2019)\citenamefont{Alekhin, Razdolski,
  Berritta, B{\"{u}}rstel, Temnov, Diesing, Bovensiepen, Woltersdorf, Oppeneer,
  and Melnikov}}]{Alekhin2019}
\bibinfo{author}{\bibfnamefont{A.}~\bibnamefont{Alekhin}},
  \bibinfo{author}{\bibfnamefont{I.}~\bibnamefont{Razdolski}},
  \bibinfo{author}{\bibfnamefont{M.}~\bibnamefont{Berritta}},
  \bibinfo{author}{\bibfnamefont{D.}~\bibnamefont{B{\"{u}}rstel}},
  \bibinfo{author}{\bibfnamefont{V.~V.} \bibnamefont{Temnov}},
  \bibinfo{author}{\bibfnamefont{D.}~\bibnamefont{Diesing}},
  \bibinfo{author}{\bibfnamefont{U.}~\bibnamefont{Bovensiepen}},
  \bibinfo{author}{\bibfnamefont{G.}~\bibnamefont{Woltersdorf}},
  \bibinfo{author}{\bibfnamefont{P.~M.} \bibnamefont{Oppeneer}},
  \bibnamefont{and} \bibinfo{author}{\bibfnamefont{A.}~\bibnamefont{Melnikov}},
  \bibinfo{journal}{Journal of Physics: Condensed Matter}
  \textbf{\bibinfo{volume}{31}}, \bibinfo{pages}{124002}
  (\bibinfo{year}{2019}),
  \urlprefix\url{https://doi.org/10.1088/1361-648X/aafd06}.

\bibitem[{\citenamefont{Thomsen et~al.}(1986)\citenamefont{Thomsen, Grahn,
  Maris, and Tauc}}]{ThomsenPRB86}
\bibinfo{author}{\bibfnamefont{C.}~\bibnamefont{Thomsen}},
  \bibinfo{author}{\bibfnamefont{H.~T.} \bibnamefont{Grahn}},
  \bibinfo{author}{\bibfnamefont{H.~J.} \bibnamefont{Maris}}, \bibnamefont{and}
  \bibinfo{author}{\bibfnamefont{J.}~\bibnamefont{Tauc}},
  \bibinfo{journal}{Phys. Rev. B} \textbf{\bibinfo{volume}{34}},
  \bibinfo{pages}{4129} (\bibinfo{year}{1986}),
  \urlprefix\url{https://link.aps.org/doi/10.1103/PhysRevB.34.4129}.

\bibitem[{\citenamefont{Temnov et~al.}(2013)\citenamefont{Temnov, Klieber,
  Nelson, Thomay, Knittel, Leitenstorfer, Makarov, Albrecht, and
  Bratschitsch}}]{TemnovNatureComm2013}
\bibinfo{author}{\bibfnamefont{V.~V.} \bibnamefont{Temnov}},
  \bibinfo{author}{\bibfnamefont{C.}~\bibnamefont{Klieber}},
  \bibinfo{author}{\bibfnamefont{K.~A.} \bibnamefont{Nelson}},
  \bibinfo{author}{\bibfnamefont{T.}~\bibnamefont{Thomay}},
  \bibinfo{author}{\bibfnamefont{V.}~\bibnamefont{Knittel}},
  \bibinfo{author}{\bibfnamefont{A.}~\bibnamefont{Leitenstorfer}},
  \bibinfo{author}{\bibfnamefont{D.}~\bibnamefont{Makarov}},
  \bibinfo{author}{\bibfnamefont{M.}~\bibnamefont{Albrecht}}, \bibnamefont{and}
  \bibinfo{author}{\bibfnamefont{R.}~\bibnamefont{Bratschitsch}},
  \bibinfo{journal}{Nature Communications} \textbf{\bibinfo{volume}{4}},
  \bibinfo{pages}{1468} (\bibinfo{year}{2013}),
  \urlprefix\url{https://doi.org/10.1038/ncomms2480}.

\bibitem[{\citenamefont{Temnov et~al.}(2016)\citenamefont{Temnov, Razdolski,
  Pezeril, Makarov, Seletskiy, Melnikov, and Nelson}}]{Temnov_2016}
\bibinfo{author}{\bibfnamefont{V.~V.} \bibnamefont{Temnov}},
  \bibinfo{author}{\bibfnamefont{I.}~\bibnamefont{Razdolski}},
  \bibinfo{author}{\bibfnamefont{T.}~\bibnamefont{Pezeril}},
  \bibinfo{author}{\bibfnamefont{D.}~\bibnamefont{Makarov}},
  \bibinfo{author}{\bibfnamefont{D.}~\bibnamefont{Seletskiy}},
  \bibinfo{author}{\bibfnamefont{A.}~\bibnamefont{Melnikov}}, \bibnamefont{and}
  \bibinfo{author}{\bibfnamefont{K.~A.} \bibnamefont{Nelson}},
  \bibinfo{journal}{Journal of Optics} \textbf{\bibinfo{volume}{18}},
  \bibinfo{pages}{093002} (\bibinfo{year}{2016}),
  \urlprefix\url{https://iopscience.iop.org/article/10.1088/1361-648X/aafd06}.

\bibitem[{\citenamefont{Farle}(1998)}]{Farle_1998}
\bibinfo{author}{\bibfnamefont{M.}~\bibnamefont{Farle}},
  \bibinfo{journal}{Reports on Progress in Physics}
  \textbf{\bibinfo{volume}{61}}, \bibinfo{pages}{755} (\bibinfo{year}{1998}),
  \urlprefix\url{https://iopscience.iop.org/article/10.1088/0034-4885/61/7/001/pdf}.

\bibitem[{\citenamefont{Salikhov et~al.}(2019)\citenamefont{Salikhov, Alekhin,
  Parpiiev, Pezeril, Makarov, Abrudan, Meckenstock, Radu, Farle, Zabel
  et~al.}}]{Salikhov2019}
\bibinfo{author}{\bibfnamefont{R.}~\bibnamefont{Salikhov}},
  \bibinfo{author}{\bibfnamefont{A.}~\bibnamefont{Alekhin}},
  \bibinfo{author}{\bibfnamefont{T.}~\bibnamefont{Parpiiev}},
  \bibinfo{author}{\bibfnamefont{T.}~\bibnamefont{Pezeril}},
  \bibinfo{author}{\bibfnamefont{D.}~\bibnamefont{Makarov}},
  \bibinfo{author}{\bibfnamefont{R.}~\bibnamefont{Abrudan}},
  \bibinfo{author}{\bibfnamefont{R.}~\bibnamefont{Meckenstock}},
  \bibinfo{author}{\bibfnamefont{F.}~\bibnamefont{Radu}},
  \bibinfo{author}{\bibfnamefont{M.}~\bibnamefont{Farle}},
  \bibinfo{author}{\bibfnamefont{H.}~\bibnamefont{Zabel}},
  \bibnamefont{et~al.}, \bibinfo{journal}{Phys. Rev. B}
  \textbf{\bibinfo{volume}{99}}, \bibinfo{pages}{104412}
  (\bibinfo{year}{2019}),
  \urlprefix\url{https://link.aps.org/doi/10.1103/PhysRevB.99.104412}.

\bibitem[{\citenamefont{Getzlaff}(2008)}]{getzlaff2008}
\bibinfo{author}{\bibfnamefont{M.}~\bibnamefont{Getzlaff}},
  \emph{\bibinfo{title}{Fundamentals of Magnetism}}
  (\bibinfo{publisher}{Springer}, \bibinfo{year}{2008}).

\bibitem[{\citenamefont{Manke et~al.}(2013)\citenamefont{Manke, Maznev,
  Klieber, Shalagatskyi, Temnov, Makarov, Baek, Eom, and
  Nelson}}]{MankeAPL2013}
\bibinfo{author}{\bibfnamefont{K.~J.} \bibnamefont{Manke}},
  \bibinfo{author}{\bibfnamefont{A.~A.} \bibnamefont{Maznev}},
  \bibinfo{author}{\bibfnamefont{C.}~\bibnamefont{Klieber}},
  \bibinfo{author}{\bibfnamefont{V.}~\bibnamefont{Shalagatskyi}},
  \bibinfo{author}{\bibfnamefont{V.~V.} \bibnamefont{Temnov}},
  \bibinfo{author}{\bibfnamefont{D.}~\bibnamefont{Makarov}},
  \bibinfo{author}{\bibfnamefont{S.-H.} \bibnamefont{Baek}},
  \bibinfo{author}{\bibfnamefont{C.-B.} \bibnamefont{Eom}}, \bibnamefont{and}
  \bibinfo{author}{\bibfnamefont{K.~A.} \bibnamefont{Nelson}},
  \bibinfo{journal}{Applied Physics Letters} \textbf{\bibinfo{volume}{103}},
  \bibinfo{pages}{173104} (\bibinfo{year}{2013}),
  \urlprefix\url{https://doi.org/10.1063/1.4826210}.

\bibitem[{\citenamefont{van Capel et~al.}(2015)\citenamefont{van Capel,
  P\'{e}ronne, and Dijkhuis}}]{vanCapelUltrasonics2015}
\bibinfo{author}{\bibfnamefont{P.~J.~S.} \bibnamefont{van Capel}},
  \bibinfo{author}{\bibfnamefont{E.}~\bibnamefont{P\'{e}ronne}},
  \bibnamefont{and} \bibinfo{author}{\bibfnamefont{J.}~\bibnamefont{Dijkhuis}},
  \bibinfo{journal}{Ultrasonics} \textbf{\bibinfo{volume}{56}},
  \bibinfo{pages}{36} (\bibinfo{year}{2015}),
  \urlprefix\url{http://www.sciencedirect.com/science/article/pii/S0041624X14002868}.

\bibitem[{\citenamefont{Kovalenko et~al.}(2013)\citenamefont{Kovalenko,
  Pezeril, and Temnov}}]{KovalenkoPRL2013}
\bibinfo{author}{\bibfnamefont{O.}~\bibnamefont{Kovalenko}},
  \bibinfo{author}{\bibfnamefont{T.}~\bibnamefont{Pezeril}}, \bibnamefont{and}
  \bibinfo{author}{\bibfnamefont{V.~V.} \bibnamefont{Temnov}},
  \bibinfo{journal}{Phys. Rev. Lett.} \textbf{\bibinfo{volume}{110}},
  \bibinfo{pages}{266602} (\bibinfo{year}{2013}),
  \urlprefix\url{https://link.aps.org/doi/10.1103/PhysRevLett.110.266602}.

\end{thebibliography}
	
\end{document}